\documentclass[aps,twocolumn,showpacs,preprintnumbers,amsmath,amssymb,nofootinbib,superscriptaddress,showkeys]{revtex4}

\usepackage{epsfig}
\usepackage{graphicx}

\begin{document}

\title{Pion Deuteron Scattering and Chiral Expansions}
\author{M. Pav\'on Valderrama}
\email{mpavon@ugr.es}

\author{E. Ruiz  Arriola}
\email{earriola@ugr.es} 

\affiliation{ Departamento de F\'{\i}sica At\'omica, Molecular y
Nuclear, Universidad de Granada, E-18071 Granada, Spain.}

\date{\today}

\begin{abstract} 
\rule{0ex}{3ex} We discuss convergence issues as well as short
distance constraints on the deuteron wave functions based on chiral
perturbation theory relevant to pion deuteron scattering.
Non-analytical terms arise in the multiple scattering series for the
pion-deuteron scattering length limiting the accuracy of the
calculations. This result resembles similar findings in the structure
of the NN interaction. The effects are found not to be numerically
large. The iso-scalar $\pi N $ scattering length from the iso-vector one
and the $\pi d$ scattering length yields values compatible with the
experimental number and with much smaller errors.
\end{abstract}

\pacs{03.65.Nk,11.10.Gh,13.75.Cs,21.30.Fe,21.45.+v}

\keywords{pion-deuteron scattering; chiral symmetries; nuclear forces; multiple scattering; perturbation theory}

\maketitle



\section{Introduction} 
\label{sec:intro}

The relevance of pion dynamics in low energy hadronic and nuclear
processes can hardly be exaggerated~\cite{Ericson:1988gk}. Low energy
theorems based on chiral symmetry provide a quantitative and model
independent insight in low energy reactions involving pions and
photons and nucleons due to the clear scale separation. The bound
state character of nuclei makes such a description more complex from a
theoretical viewpoint because many relevant scales enter into the
problem, but one expects simplifications to occur in the limit of weak
binding and low energies as it is the case for the deuteron. Actually,
the possibility of computing Pion-Deuteron scattering (for a review
see e.g.~\cite{Thomas:1979xu} and references therein) in a model
independent way was one of the original motivations to introduce
Effective Field Theory (EFT) approaches~\cite{Weinberg:1990rz} based
on the chiral symmetry of QCD and to derive the corresponding low
energy theorem~\cite{Weinberg:1992yk}. This approach has been
vigorously extended in the last decade to the calculation of low
energy reactions for finite nuclei (for comprehensive reviews see
e.g. Ref.~\cite{Bedaque:2002mn,Phillips:2002da,Phillips:2005vv}).
This is in fact a rather complicated process since there are many
corrections and physical effects which add up to the final
result~\cite{Ericson:2000md,Doring:2004kt} and it is a challenge for
chiral approaches to make reliably calculations based on a priori
estimates of the accuracy. The use of the original power counting
suggested by Weinberg to extract the $\pi-$neutron scattering length from
the known $\pi-$proton and $\pi-$deuteron ones was suggested in
Ref.~\cite{Beane:1997yg}. This power counting has been
modified in Ref.~\cite{Beane:2002wk} to account for the long distance
enhancement since the deuteron wavefunctions extend far beyond the
pion Compton wavelength. Their result resembles the single and double
scattering terms of the widely used older
approaches~\cite{Thomas:1979xu,Ericson:2000md,Doring:2004kt} where the
deuteron inverse moments, $\langle r^{-n}\rangle $, play an essential
role. Re-scattering, deuteron recoil and binding effects can be
re-summed to all orders in a rather elegant
formula~\cite{Deloff:2001zp} in the isospin limit (for a review see
e.g. Ref.~\cite{Deloff:2003ns}). Nucleon recoil effects have also been
shown to be small~\cite{Baru:2004kw}.  Pionic deuterium has been discussed in 
Ref.~\cite{Meissner:2005bz}. Traditionally, these matrix
elements have been evaluated using potential model calculations for
the deuteron wave functions, which generally produce divergent results
for $ \langle r^{-n} \rangle$ (with $n > 2$). Perturbative wave
functions based on OPE potentials yield divergences already for
$\langle r^{-1} \rangle $~\cite{Borasoy:2003gf,PavonValderrama:2005gu}
and only recently  has it been realized that non-perturbative
calculations have a quite different convergent
behaviour~\cite{PavonValderrama:2005gu,Nogga:2005fv,Platter:2006pt}.

Following insightful previous works~\cite{Martorell94,Beane:2001bc},
we have developed in a series of
papers~\cite{PavonValderrama:2003np,PavonValderrama:2004nb,
PavonValderrama:2005gu,Valderrama:2005wv,PavonValderrama:2005uj} a
framework for the treatment of the NN interaction in a model
independent way based on long distance correlations among physical
observables. In our approach the long distance chiral NN- One Pion
Exchange (OPE) and Two Pion Exchange (TPE) potentials computed within
perturbation
theory~\cite{Kaiser:1997mw,Friar:1999sj,Rentmeester:1999vw}, are
iterated to all orders in the Schr\"odinger equation very much in the
spirit of the original Weinberg approach~\cite{Weinberg:1990rz}
although some additional subtleties are
encountered~\cite{PavonValderrama:2004nb,Nogga:2005hy,Valderrama:2005wv}. Due
to the power like singular character of the chiral potentials at the
origin, a radial short distance cut-off is imposed which ultimately is
removed by renormalization fixing some low energy observables or
deuteron properties.  This procedure turns out not to correspond to
the imposition of a strict power counting in the sense that
corrections to physical observables might be estimated {\it a priori}
by dimensional arguments, and hence departs from the original EFT
program. Instead, one can show that naive perturbative expansions of
renormalized scattering amplitudes do indeed display divergences which
can be traced to given non-analyticities in the expansion parameter of
the solution with the long distance potential~\cite{Valderrama:2005wv}
. More specifically, if the long distance potential is written as
$V(r)=V_{\rm OPE} (r) + \lambda V_{\rm TPE}(r) + \dots $ one has that
the scattering amplitude at low energies fulfills ${\cal A} - {\cal
A}_{\rm OPE} = {\cal O} ( \lambda^\alpha) $ with $0 < \alpha < 1$ for
singular potentials which fulfill the condition that $ V_{\rm OPE} (r)
\gg V_{\rm TPE}(r) $ at large distances. The value of the exponent
$\alpha$ depends on the particular structure of the potentials
involved. This weird fractional power counting is uncomfortable from a
theoretical viewpoint and admittedly non conventional but does not
contradict the physical requirement that effects which are small at
large distances become parametrically small at low energies. This
happens to be so even if at short distances both singular potentials
fulfill the opposite relation $ V_{\rm OPE} (r) \ll V_{\rm TPE}(r)
$. As a consequence standard perturbation theory cannot be applied.
The singularity of the potentials just provides a non-analytical
enhancement of the perturbation. Thus, there is a lack of
systematics a priori but one can always check the corrections to be
small numerically as it turns out to be the case for the OPE and TPE
potentials. For long distance potentials, which are local, as it is
the case for the OPE and TPE
potentials~\cite{Kaiser:1997mw,Friar:1999sj,Rentmeester:1999vw},
severe restrictions on the possible counterterms are derived from the
physical requirement that the wave function be small in the short
distance and unknown region as the cut-off is
removed~\cite{Valderrama:2005wv}. Calculations based on finite
cut-offs do not show these problems but then the model independence is
manifestly lost since the particular regularization acts itself as a
short distance model.

With these new insights in mind we want to address in this paper the
issue of $\pi d$ scattering and, more specifically, the calculation of
the corresponding scattering length within the multiple scattering
formalism to all orders as suggested by Deloff~\cite{Deloff:2001zp}
where, as already mentioned, the inverse moments of the deuteron wave
function play a role. In our OPE analysis of the
deuteron~\cite{PavonValderrama:2005gu} we showed that after
renormalization the first inverse moment is finite due to the peculiar
short distance behaviour of the deuteron wave function and {\it
because} the potential at short distances presents a $1/r^3$
singularity. The analytical structure at short distances was
determined to high orders.  A similar observation was made by Nogga
and Hanhart Ref.~\cite{Nogga:2005fv} using the momentum space wave
functions of Ref.~\cite{Nogga:2005hy}. Recently, Platter and
Phillips~\cite{Platter:2006pt} have analyzed this matrix element as
well as the second inverse moment and have shown that a direct
treatment in coordinate space~\cite{PavonValderrama:2005gu} allows
naturally for a rather clean extrapolation of the renormalized matrix
element.

In the present work we discuss and extend these developments to the
TPE case on the light of the multiple scattering formalism to all
orders (Sect.~\ref{sec:multi}). Let us remind that the truncation of
such an expansion where the coefficients are in fact the inverse
deuteron moments is motivated by the smallness of the $\pi N$
scattering lengths. We show that higher order inverse moments are
indeed convergent when the deuteron wave functions are built by fully
iterating the long distance potentials (See
Sect.~\ref{sec:short}). The finite values are computed in
Sect.~\ref{sec:inverse}.  However, to any given approximation of the
long distance potential there is always a finite value of the order of
the moment above which none of them converges. In contrast, the
re-summation of re-scattering effects provides convergent results and
indeed exhibit a logarithmic enhancement of the multiple scattering
expansion parameter (Sect.~\ref{sec:non-anal}). We use these
approaches to examine the determination of the iso-scalar $\pi N$
scattering length within several schemes. Boost and finite range
corrections are discussed qualitatively in
Sect.~\ref{sec:boost}. Finally, in Sect.~\ref{sec:concl} we come to
the conclusions.

\section{The multiple scattering expansion and short distance sensitivity}
\label{sec:multi}

In a remarkable paper Deloff~\cite{Deloff:2001zp} has analyzed $\pi
d$ scattering at threshold by solving the Faddeev equation and has
worked out the scattering length within a zero range model based on a
boundary condition for the relative pion-nucleon wave function. In
field operator language this corresponds to a contact $NN\pi\pi$
interaction, which can be regulated in this context by a radial
regulator. In this approximation, intrinsic finite size effects of the
$\pi N$ interaction are not included (see
e.g.~\cite{Ericson:2000md,Doring:2004kt} for a recent discussion on
these and other effects).

The result he found in Ref.~\cite{Deloff:2001zp} for the scattering
length is quite simple
\begin{eqnarray}
a_{\pi d} = \frac{2}{1+m/2M}\int_0^\infty dr \left[ u (r)^2+
w (r)^2 \right] A_{\pi d} (r)
\label{eq:a_pi-d}
\end{eqnarray} 
where $u (r)$ and $w (r)$ are  S- and D-wave deuteron
wave functions satisfying the coupled channel $^3S_1 - ^3D_1 $ set of
equations 
\begin{eqnarray}
-u '' (r) + U_{s} (r) u (r) + U_{sd} (r) w (r) &=& -\gamma^2 u
 (r) \, ,\nonumber  \\ -w '' (r) + U_{sd} (r) u (r) + \left[U_{d} (r) +
 \frac{6}{r^2} \right] w (r) &=& -\gamma^2 w (r) \, , \nonumber \\
\label{eq:sch_coupled} 
\end{eqnarray}
and with asymptotic conditions 
\begin{eqnarray}
u (r) &\to & A_S e^{-\gamma r} \, , \nonumber \\ w (r) & \to & A_D
e^{-\gamma r} \left( 1 + \frac{3}{\gamma r} + \frac{3}{(\gamma r)^2}
\right) \, ,
\label{eq:bcinfty_coupled} 
\end{eqnarray}
where $ \gamma = \sqrt{M B} = 0.231605\,{\rm fm}^{-1}$ is the deuteron 
wave number, $A_S$ is the normalization factor such that 
\begin{eqnarray}
\int_0^\infty dr \left[ u(r)^2+ w(r)^2 \right] = 1  \, , 
\end{eqnarray} 
and the asymptotic D/S ratio parameter is defined by $\eta=A_D/A_S$.
For conventions and numerical values of parameters we use
Ref.~\cite{PavonValderrama:2005gu,Valderrama:2005wv} throughout the
paper.  

\begin{figure}[]
\begin{center}
\epsfig{figure=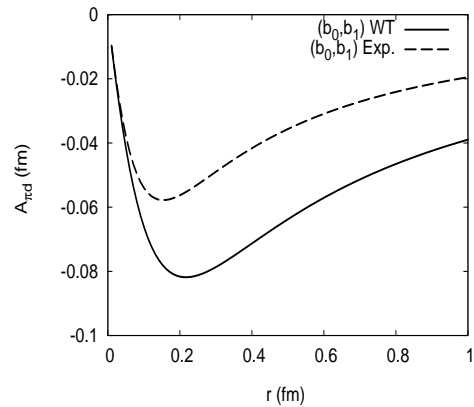,height=5.5cm,width=6.5cm}
\end{center}
\caption{The Deloff function $A_{\pi d} (r) $ (in fm) as a function of
distance in (fm). We use the Weinberg-Tomozawa values as well as the
experimental ones for the s-wave $\pi N$ scattering lengths and
include recoil and binding corrections.  }
\label{fig:Deloff}
\end{figure}

The function $A_{\pi d} (r) $ in Eq.~(\ref{eq:a_pi-d}), which will be
called Deloff function for short, is given by
\begin{eqnarray}
A_{\pi d} (r) = \frac{\tilde{b}_0 + 
(\tilde{b}_0+\tilde{b}_1)(\tilde{b}_0-2 \tilde{b}_1) / r }{1 - \tilde{b}_1/r
  -(\tilde{b}_0+\tilde{b}_1)(\tilde{b}_0-2 \tilde{b}_1)/r^2} \, , 
\end{eqnarray}
with $\tilde{b}_i = (1 + m/M) b_i$,
being $b_0$ and $b_1$ the $\pi N$ scattering lengths according to the
standard decomposition (assuming isospin symmetry)
\begin{eqnarray} 
{\cal F}_{\pi N} = b_0 + b_1 \vec t \cdot \vec \tau \, . 
\end{eqnarray} 
Recoil and deuteron binding effects are taken into account by making
the simple replacements~\cite{Deloff:2001zp}
\begin{eqnarray}\label{eq:RB-corr}
\tilde{b}_i \to \hat{b}_i = \frac{(1 + m/M)}{1+\kappa b_i} b_i \qquad , 
\quad 1/r \to e^{-\kappa r } / r \, . 
\end{eqnarray}
where $\kappa = \gamma \sqrt{2m / (m + M)} = 0.117261\,{\rm fm}^{-1}$.
The previous Eq.~(\ref{eq:a_pi-d}) sums up all multiple scattering
effects due to zero range $\pi N $ interactions. It does not include
Fermi motion, higher partial waves contributions nor finite range $\pi
N $ corrections. We will comment on these corrections at the end of
this paper. When expanded for small $b_0$ and $b_1$ one gets the
result
\begin{eqnarray}
 a_{\pi d} = \frac{2}{1+\frac{m}{2M}} &\Big[& b_0 + (b_0^2 - 2 b_1^2)
 \Big\langle \frac1{r} \Big\rangle \nonumber \\ &+& (b_0^3 - 2 b_1^2
 b_0 - 2b_1^3 ) \Big\langle \frac1{r^2} \Big\rangle \nonumber \\ &+&
 (b_0^4 - 4 b_1^2 b_0^2 + 2 b_1^4 ) \Big\langle \frac1{r^3} 
 \Big\rangle + \dots  \Big] 
 \label{eq:multiple} 
\end{eqnarray} 
which has been used quite often truncated to second
order~\cite{Ericson:1988gk}. Here, the deuteron wave function average
is defined as
\begin{eqnarray}
\Big\langle \frac1{r^n} \Big\rangle = 
\int_0^\infty dr \frac{ u (r)^2+
w (r)^2 }{r^n}
\end{eqnarray}
The multiple scattering expansion is motivated by the smallness of the
$s-$wave $\pi N $ scattering lengths.  Actually, such an expansion
finds theoretical support from Chiral Perturbation Theory since at
lowest order one has the Weinberg-Tomozawa (WT)
relations~\cite{Ericson:1988gk}
\begin{eqnarray}
b_0^W &=& \frac13 \left( a_1 + 2 a_3 \right) =0 \\ 
b_1^W &=& \frac13 \left( a_3 - a_1 \right) = - 
\frac{m_\pi}{8 \pi ( 1+ m_\pi /M ) f_\pi^2 } \, , 
\end{eqnarray} 
which compare rather well with the experimental numbers extracted from
pionic hydrogen~\cite{Schroder:1999uq}
\begin{eqnarray}
b_0 &=& -(0.22 \pm 0.43) \times 10^{-2} m_\pi^{-1}  \\ 
b_1 &=& -(9.05 \pm 0.42) \times 10^{-2} m_\pi^{-1}  \, .
\end{eqnarray} 
Higher order corrections to the current algebra relations have been
computed via standard ChPT methods~\cite{Fettes:2001cr}. A picture of
the Deloff function can be seen in Fig.~\ref{fig:Deloff} for the WT
values as well as the previous values from
Ref.~\cite{Schroder:1999uq}.  On the other hand, the measured $\pi^- d
$ scattering length is~\cite{Schroder:2001rc,Hauser:1998yd}
\begin{eqnarray}
a_{\pi^- d} = \left[ -252 \pm 5 ({\rm stat.}) \pm 5 ({\rm syst.})
\right] \times 10^{-4} m_\pi^{-1}  \, .
\end{eqnarray} 
From the viewpoint of the Chiral nuclear approach, the issue that we
address here is to determine the compatibility of all scattering
lengths using both the multiple scattering expansion as well as the
deuteron wave functions based on chiral potentials. 

\medskip

In the case of the zero range $\pi N $ interaction the convergence of
the multiple scattering expansion, Eq.~(\ref{eq:multiple}), would
require in particular that any of the inverse moments of the deuteron
wave function must be finite at the origin. However, potential model
wave functions based on regular potentials, i.e. $r^2 U(r) \to 0 $ for
$r \to 0$, are dominated by the centrifugal term at short distances
and hence satisfy the regularity conditions at the origin,
\begin{eqnarray}
u (r) \sim r \qquad , \, w (r) \sim r^5  \, ,  
\end{eqnarray} 
reflecting their $L=0$ and $L=2$ angular momentum character
respectively. So, it is clear that negative moments fail to converge
starting at third order where a logarithmic divergence takes place. In
addition, the short distance behaviour of the deuteron wave function
becomes relevant for $\langle r^{-2} \rangle$ and, in fact, potential
models indeed exhibit this sensitivity. The third and higher inverse
moments are divergent. This strong short distance dependence looks
very weird and counter intuitive, since we are looking at
pion-deuteron scattering at zero energy, where the wavelength of the
incoming and outgoing pion is much larger than any of the other length
scales of the problem. So, we regard this effect as a mathematical
artifact of the expansion and the potential model wave functions, and
not as a genuine physical feature.  It certainly does not agree with
the philosophy underlying Effective Field Theories, namely that low
energy physics does not depend on short distance details. On the other
hand the full formula does not present this problem because in spite
of going to a finite limit at long distances, $A_{\pi d} (r) \to b_0
$, there is a linear short distance suppression of the operator,
$A_{\pi d} (r)\sim - r $, as a result of the re-summation and in
agreement with the EFT expectations. These considerations suggest that
there may be problems with the convergence of multiple scattering
expansion, which is ultimately motivated by the weak s-wave $\pi N$
interaction at threshold and which fits quite naturally within Chiral
Perturbation Theory. We will show below that the problem is related to
the implicit assumption of analyticity in the $\pi N $ scattering
lengths.

\section{Short distance constraints on matrix elements and Singular potentials}
\label{sec:short} 

One relevant question is whether the multiple scattering expansion so
widely used can still be undertaken and to what order in the case of
the zero range $\pi N $ interaction. Obviously, for this to happen at
a given finite order, say $\langle r^{-k} \rangle < \infty$, we must
have
\begin{eqnarray}
u (r) \sim r^{(k-1)/2 \,+\,\epsilon} 
\qquad , \, 
w (r) \sim r^{(k-1)/2 \,+\, \epsilon}  \, ,   
\end{eqnarray}
where $\epsilon > 0$, and which implicitly requires that short
distances cannot be dominated by the centrifugal term. The only way
how this may happen is that the potential becomes more singular than
the centrifugal barrier, which behaves as $1/r^2$. Indeed for a
potential which diverges like a given power $U(r) \sim r^{-2 n}$ the
wave function has the power behaviour $u(r) \sim r^{n/2} $ (up to some
exponential or oscillatory function depending on the attractive or
repulsive character of the potential at short distances)~\footnote{In
the limit $n \to \infty$ this includes also the possibility of an
infiniteley repulsive hard core potential.}. Thus, the maximal value
for which the inverse moments converge fulfill
\begin{eqnarray}
\langle r^{-k} \rangle < \infty \qquad {\rm if} \qquad U(r) \sim
r^{-2k - 2 \,+\, \epsilon} \, . 
\end{eqnarray} 
It is remarkable that chiral potentials do indeed exhibit these short
distance singularities required by finiteness on the inverse moments
of the deuteron wave function. Actually, a chiral expansion of the
potential reads~\cite{Kaiser:1997mw,Friar:1999sj,Rentmeester:1999vw},
\begin{eqnarray}
U(r) &=& \frac{M m^3}{f^2} F^{(0)} (mr) + \frac{M m^5 }{f^4} F^{(2)}
(mr) \nonumber\\ 
&& + \frac{m^6}{f^4} F^{(3)} (mr) + \dots 
\end{eqnarray} 
One can rewrite the expansion as
\begin{eqnarray}
U(r) &=& \frac{M }{f^2 r^3 } G^{(0)} (mr) +
\frac{M }{f^4 r^5 } G^{(2)} (mr) \nonumber \\ &&
\frac{1}{f^4 r^6 } G^{(3)} (mr) + \dots 
\end{eqnarray} 
where the functions $G^{(n)} (mr)$ have a finite limit for vanishing
argument. Thus in the short distance limit 
\begin{eqnarray}
U(r) & \to & \frac{M }{f^2 r^3 } G^{(0)} (0) +
\frac{M }{f^4 r^5 } G^{(2)} (0) \nonumber \\ &&
\frac{1}{f^4 r^6 } G^{(3)} (0) + \dots 
\end{eqnarray} 
Note that in this limit the pion mass dependence
disappears~\footnote{The short distance however {\it does not}
coincide with the chiral limit; in the latter case less singular
subleading powers are obtained.}. In the deuteron case the
coefficients in the former expressions become
matrices~\cite{Valderrama:2005wv} (see Appendix~\ref{sec:app}).

Thus, in the absence of the long distance potential none of the
inverse moments is finite, while at LO the moments $\langle r^{-1}
\rangle$ and $\langle r^{-2} \rangle$ are finite, and at NLO and
NNLO $\langle r^{-1} \rangle $,  $\langle r^{-2} \rangle $ and
$\langle r^{-3} \rangle $ are also finite.

Given the fact that the short distance behaviour of the chiral
potentials does not depend on the pion mass, the short distance
contribution to the inverse moments is dominated by the corresponding
short distance scale, $R$, so one has
\begin{eqnarray}
\Big\langle r^{-k} \Big\rangle_{\rm short} \sim R^{-k} \, , 
\end{eqnarray}  
provided the integral is convergent. Thus, if we use the OPE exchange
potential, $R \sim M/f^2 $ we see that inverse moments do indeed
become large in the limit $R \to 0$. As shown in our previous work on
the deuteron~\cite{PavonValderrama:2005gu}, long distance perturbation
theory mistreats the behaviour of the wave function at short
distances, introducing a very unnatural strong short distance
dependence. Actually the first order contribution to the deuteron wave
function diverges. As a consequence, in the OPE potential, $\langle
r^{-1} \rangle $ also diverges as first noted in
Ref.~\cite{Borasoy:2003gf} using the PDS subtraction scheme. This was
the reason to choose always the regular solutions of the fully
iterated potential at the origin for the OPE
case~\cite{PavonValderrama:2005gu}. The divergence persists also when
the TPE exchange potential is treated in perturbation theory on the
OPE distorted wave basis as a zeroth order approximation, since as
pointed out in Ref.~\cite{Valderrama:2005wv}, the deuteron
perturbative wave functions diverges strongly at the origin.

As we see, the multiple scattering expansion of the pion-deuteron
scattering length involves negative moments which become more
convergent and hence more insensitive to short distance details when
the NN potential is more accurately described at shorter
distances. For this finiteness of negative moments to occur it is
essential that the singular potentials be {\it fully iterated} since
the short distance behaviour is highly non perturbative. This gives us
some confidence on the treatment of the singular potentials and the
renormalization process adopted in our previous
works~\cite{PavonValderrama:2005gu,Valderrama:2005wv}. As we have also
stressed in the introduction, the successive improvements on the
potential are {\it parametrically small} although in a non-analytical
way for the low energy NN observables.  Note that this is {\it not}
the case for the inverse moments.  Nevertheless, we will see that a
similar non-analytical behaviour occurs in the perturbative treatment
of the pion-deuteron scattering length via a multiple scattering
expansion.

\section{Inverse moments for fully iterated Chiral Potentials }
\label{sec:inverse}

The calculation of the inverse moments for the deuteron wave functions
for the OPE and TPE potentials is in principle straightforward.  In
the OPE case the first inverse moment was calculated in
Ref.~\cite{PavonValderrama:2005gu} and then in
Ref.~\cite{Nogga:2005fv}. These numbers have been checked in
Ref.~\cite{Platter:2006pt}. In the later reference also the second
inverse moment $\langle r^{-2} \rangle $ has been estimated. 
Here we confirm and extend these results to the first three
finite inverse moments in the TPE case.

\begin{figure}[]
\begin{center}
\epsfig{figure=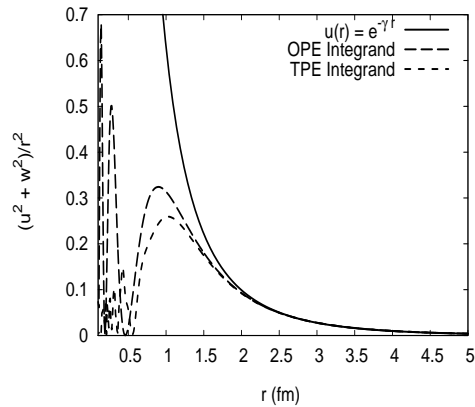,height=5.5cm,width=6.5cm}
\end{center}
\caption{The integrand $(u^2 + w^2)/r^2$ for the pion-less, OPE and TPE
deuteron wave functions.}
\label{fig:integrand.ir2}
\end{figure}

One technical aspect to consider in the present calculation
corresponds to the short distance contribution of the matrix elements.
As an illustration we plot in Fig.~\ref{fig:integrand.ir2} the
integrand corresponding to the inverse square moment, $(u^2+w^2)/r^2$
for both the OPE and the TPE potentials. As we see, a substantial
contribution to the integral is dominated by the short distance region
making the convergence at short distances numerically unreliable. In
the OPE case this can be fixed by using the analytical solutions found
in our previous work~\cite{PavonValderrama:2005gu} and computing the
integral analytically in the short distance region. In the
appendix~\ref{sec:app} we analyze the problem for the OPE and TPE in
more detail. Actually, the possibility of making these estimates
analytically as well as the determination of the convergence of matrix
elements is a virtue of the coordinate space method which exploits the
locality of the chiral potentials in a natural way~\footnote{This is
in contrast to momentum space methods where computer space limitations
may suggest a seeming convergence on the momentum space cut-off but
still far from the infinite cut-off limit result (see e.g. Fig.5 in
Ref.  \cite{Platter:2006pt}).}. One important feature is that $20 \%$
of the total value comes from the region below $0.2 {\rm fm}$. Up to
tiny oscillations the convergence behaves as $\sqrt{r_c}$ for $r_c \to
0$.  

Following
Ref.~\cite{PavonValderrama:2005gu} we use the superposition principle
of boundary conditions and write the deuteron wave functions as
\begin{eqnarray}
u (r) &=& u_S (r) + \eta \, u_D (r) \, , \nonumber \\ w (r) &=& w_S
(r) + \eta \, w_D (r) \, . 
\label{eq:sup_bound} 
\end{eqnarray}
For the OPE potential the short distance bevaviour has been displayed
in Ref.~\cite{PavonValderrama:2005gu}. The regularity condition at the
origin fixes $\eta_{\rm OPE} = 0.02633$. In the TPE potential we refer
to Ref.~\cite{Valderrama:2005wv} for the short distance behaviour. In
such a case $\eta$ is a free parameter which can be fixed to the
experimental value with the corresponding uncertainties $\eta_{\rm
exp}=0.0256(4)$.  Using Eqs.~(\ref{eq:sup_bound}) we can display the
$\eta$ parameter dependence explicitly in observables involving
deuteron wave functions,
\begin{eqnarray}
\Big\langle \frac1{r^n} \Big\rangle_{\rm TPE} = \frac{ \Big\langle \frac1{r^n}
\Big\rangle_{SS} + 2 \eta \Big\langle \frac1{r^n} \Big\rangle_{SD} +
\eta^2 \Big\langle \frac1{r^n} \Big\rangle_{DD}}{ 1_{SS} + 2 \eta
1_{SD} + \eta^2 1_{DD} }
\end{eqnarray} 
Numerically, for the chiral constants deduced in Ref.~\cite{Entem:2003ft} (Set IV in our work) we find the expressions, 
\begin{eqnarray}
\Big\langle \frac1r \Big\rangle_{\rm TPE} &=& \frac{2.43214 - 181.998\,\eta +
  3942.\,{\eta}^2}{3.39582 - 189.582\,\eta + 4175.2\,{\eta}^2}  \nonumber \\
\Big\langle \frac1{r^2} \Big\rangle_{\rm TPE} &=& \frac{2.52146 - 202.658\,\eta
  + 4350.4\,{\eta}^2}{3.39582 - 189.582\,\eta + 4175.2\,{\eta}^2} \nonumber \\
\Big\langle \frac1{r^3} \Big\rangle_{\rm TPE} &=& \frac{3.19633 - 261.897\,\eta
  + 5575.01\,{\eta}^2}{3.39582 - 189.582\,\eta + 4175.2\,{\eta}^2} \nonumber \\ \end{eqnarray} 
And also 
\begin{eqnarray}
\Big\langle \frac{e^{-\kappa r}}{r} \Big\rangle_{\rm TPE} &=&
\frac{3.08822 - 221.425\,\eta + 4807.6\,{\eta}^2}{3.39582 -
  189.582\,\eta + 4175.2\,{\eta}^2} \nonumber \\ \Big\langle
\frac{e^{-2\kappa r}}{r^2} \Big\rangle_{\rm TPE} &=& \frac{3.66165 -
  287.55\,\eta + 6189.54\,{\eta}^2}{3.39582 - 189.582\,\eta +
  4175.2\,{\eta}^2} \nonumber \\ \Big\langle \frac{e^{-3\kappa
    r}}{r^3} \Big\rangle_{\rm TPE} &=& \frac{5.14335 - 417.313\,\eta +
  8914.07\,{\eta}^2}{3.39582 - 189.582\,\eta + 4175.2\,{\eta}^2} \nonumber \\ 
\end{eqnarray} 
The numerical coefficients depend {\it solely} on the TPE potential, $\gamma$
and $\kappa$ for which we take the values $0.231605$ and 
$0.117261 {\rm fm}^{-1}$ respectively. Our
results for the computed moments are summarized in table~\ref{tab:inverse_r_moments}. Errors on the short distance cut-off
scale linearly, quadratically and cubically for $\langle r^{-3}
\rangle $, $\langle r^{-2} \rangle $ and $\langle r^{-1} \rangle $
respectively (see appendix~\ref{sec:app}).

\medskip

\begin{table*}
\begin{center}
\begin{tabular}{|c|c|c|c|c|c|c|c|c|c|}
\hline \hline
& $\gamma\,({\rm fm}^{-1})$ & $\eta$ 
& $\langle \frac{1}{r} \rangle \, ({\rm fm}^{-1})$ 
& $\langle \frac{1}{r^2} \rangle \, ({\rm fm}^{-2})$ 
& $\langle \frac{1}{r^3} \rangle \, ({\rm fm}^{-3})$ 
& $\langle \frac{e^{-\kappa r}}{r} \rangle \, ({\rm fm}^{-1})$ 
& $\langle \frac{e^{-2 \kappa r}}{r^2} \rangle \, ({\rm fm}^{-2})$ 
& $\langle \frac{e^{-3 \kappa r}}{r^3} \rangle \, ({\rm fm}^{-3})$ 
\\ 
\hline
{\rm $u(r) = e^{-\gamma r}$} & Input & 0 & $\infty$ & $\infty$ & $\infty$ & $\infty$ & $\infty$ & $\infty$ \\ 
\hline
{\rm OPE} & Input & 0.02633 & 0.4782(7) & 0.4208(10) & $\infty$ 
& 0.380(1) & 0.3313(6) & $\infty$ \\ 
{\rm OPE}$^*$ & Input & 0.02555 & 0.4861(10) & 0.434(3) & $\infty$ 
& 0.387(1) & 0.343(4) & $\infty$ \\ 
\hline 
{\rm TPE-Set I} & Input & Input & 0.424(3) & 0.248(3) & 0.213(5) 
& 0.326(3) & 0.170(4) & 0.146(5) \\ 
{\rm TPE-Set II} & Input &  Input & 0.424(4) & 0.248(4) & 0.214(6) 
& 0.327(3) & 0.171(4) & 0.147(5) \\
{\rm TPE-Set III} & Input & Input & 0.436(3) & 0.265(6) & 0.239(10) 
& 0.338(3) & 0.185(5) & 0.167(9)
\\ 
{\rm TPE-Set IV} & Input & Input & 0.447(5) & 0.284(8) & 0.276(13) 
& 0.349(5) & 0.202(7) & 0.198(12)
\\ 
\hline
NijmII & 0.231605 & 0.02521 & 0.4502 & 0.2868 & $\infty$ 
& 0.3519 & 0.2032 & $\infty$ \\
Reid93 & 0.231605 & 0.02514 & 0.4515 & 0.2924 & $\infty$ 
& 0.3531 & 0.2084 & $\infty$\\
\hline
Exp & 0.231605 & 0.0256(4) & - & - & - & - & - & -
\\
\hline 
\hline
\end{tabular}
\end{center}
\caption{ Inverse moments of the radius and $e^{-\kappa\,r} / r$
moments for the deuteron. We consider the OPE and TPE potentials; in
the case of the OPE potential we have taken $g_{\pi NN} = 13.08$ (OPE)
and $g_A = 1.26$(OPE$^*$), while in the TPE case we show the results
corresponding to the four set of chiral couplings considered along
this work.  In the OPE case the error is estimated by varying the
semiclassical matching radius in the range $0.1 - 0.2\,{\rm fm}$ (see
Appendix~(\ref{sec:app})), while in the TPE case the error comes from
the experimental uncertainty of the $D/S$ ratio, $\eta = 0.0256(4)$.
TPE Sets I,II,II and IV refer to the chiral parameters, $c_1$, $c_3$
and $c_4$ of Refs.~\cite{Buettiker:1999ap},
\cite{Rentmeester:1999vw},\cite{Entem:2002sf} and \cite{Entem:2003ft}
respectively.}
\label{tab:inverse_r_moments}
\end{table*}

Besides the fact that the non-perturbative inclusion of long distance
chiral potentials to all orders provides convergent inverse moments,
one of the features one observes from inspection of table
\ref{tab:inverse_r_moments} is the size of the changes induced when
going from OPE to TPE potentials.  Whereas the TPE effects have
moderate effects on the $\langle r^{-1} \rangle$ moment as compared to
the OPE value, almost a factor of two reduction for $\langle
r^{-2} \rangle$ is obtained. This is to be expected as the TPE
potential modifies the short distance region~\footnote{A similar
factor of two was also found in the effective range parameter for the
$^1S_0$ phase shift when going from OPE to TPE.}. Moreover, we get a
large reduction of the second inverse moment since $\langle r^{-2}
\rangle_{\rm OPE}= 0.41 {\rm fm}^{-2}$ while $\langle r^{-2}
\rangle_{\rm TPE}= 0.25-0.28 {\rm fm}^{-2}$ very close to the value
$0.286-0.345 {\rm fm}^{-2}$ quoted in Ref.~\cite{Beane:2002wk}. As we
also found in our previous work~\cite{Valderrama:2005wv} the Set
IV~\cite{Entem:2003ft} agrees best with the NijmII and Reid93
potential values.

Finally, let us note that, as already argued above, the more singular
the potential at short distances the more convergent the matrix
elements, and in particular short distance contributions to the TPE
potential are {\it much smaller} than those in the OPE case. For
instance for the second inverse moment in the region below $r_c=0.2
{\rm fm}$ one gets $\langle r^{-2} \rangle_{\rm OPE}^{\rm short}=0.08
{\rm fm}^{-2}$ while $\langle r^{-2} \rangle_{\rm TPE}^{\rm
short}=0.003{\rm fm}^{-2}$. This also suggests that finite cut-off
effects in matrix elements become smaller when the long distance
potential is improved at lower distances.

\section{Divergent inverse moments and Non-analytical behaviour}
\label{sec:non-anal}

As we have also discussed, the use of the Deloff function (with or
without binding and recoil) produces finite numbers regardless on the
short distance behaviour of the deuteron wave function assuming it can
be normalized. On the other hand when the multiple scattering
expansion is undertaken divergences appear at some stage. Although
this appears to be a bit puzzling, the mathematical reason why this is
happening is indeed quite simple. If we use for definiteness the WT
values for the $\pi N$ scattering lengths, the denominator has complex
conjugated poles located
\begin{eqnarray} 
r = - \frac{(1 \pm i \sqrt{7})m}{16 \pi f^2 } + \dots 
\end{eqnarray} 
where binding and recoil corrections have been neglected for
clarity.  
In this approximation~\footnote{The full expression is
$$ |r e^{\kappa r} |= \frac{\sqrt{2} m }{(1+m/M) \left[8 \pi f^2
    (1+m/M) - m \kappa \right]}
$$} the radius of convergence becomes $|r|= m / 4 \pi \sqrt{2} f^2 $
which is a rather small distance and goes to zero in the chiral limit.
For finite values of $b_1$ one can only expand above that region.  The
relevant scale is given by $ b_1 \sim -0.2 {\rm fm}$ so that in the
limit of small $b_1$ we become sensitive to the short distance
behaviour. Numerically the radius of convergence is given by
$|r_c|=0.15 {\rm fm}$, so the the large $r$ expansion converges only
for $r > |r_c| $. Thus, if we cut-off the integrand below such a value
we get a convergent expansion but at the same time an important piece
of physics is neglected.  The short distance enhancement can be seen
by displaying at the function $A_{\pi d}(r)$ in
Fig.~\ref{fig:Deloff}. The bump takes place at $r \sim 0.2 {\rm fm} $
so it is indeed true that a relevant contribution comes from short
distances.

The kind of non-analyticity that appears can be easily illustrated
with the asymptotic deuteron wave function 
\begin{eqnarray}
u (r) = \sqrt{2 \gamma} e^{- \gamma r} \qquad  , \, w (r) =0  \, ,  
\label{eq:short}
\end{eqnarray} 
which corresponds to the pion-less theory. Even though the first term
in the multiple scattering expansion in Eq.~(\ref{eq:multiple}) does
not converge, the full formula given by  Eq.~(\ref{eq:a_pi-d}) yields a finite
analytical result. Neglecting $m_\pi /M$ corrections and taking
$b_0=0$ for illustration purposes one gets
\begin{eqnarray}
a_{\pi d} = 4 b_1^2 \gamma I(- b_1 \gamma )  
\end{eqnarray} 
where
\begin{eqnarray}
I(t) &=& \int_0^\infty \frac{e^{-x t}x }{(x-x_1)(x-x_2)} \nonumber \\
&=& \frac{e^{-t x_1} x_1 \Gamma(0, -t x_1) - e^{-t x_2} x_2 \Gamma(0,
-t x_2)}{x_1-x_2}  \, . 
\end{eqnarray} 
In the case $b_0=0$ we have $x_{1,2}= (1\pm i \sqrt{7})/2 $ and
$\Gamma(0,z) = \int_z^\infty e^{-t} / t \, dt $ is the incomplete gamma
function which has a logarithmic branch cut at $z=0$. Thus, there is a
branch cut singularity at $b_1=0$, the leading term being 
\begin{eqnarray}
a_{\pi d} \sim -4 b_1^2 \gamma \log ( -b_1 \gamma ) = -3.06 \times
10^{-2} m_\pi^{-1} 
\end{eqnarray} 
This example shows explicitly the kind of enhancement that one might
expect. The full result with the short distance wave function and
using the WT $\pi N$ scattering lengths $a_{\pi d} = -2.815 \cdot 10^{-2} 
m_\pi^{-1}$, a quite reasonable value taking into account the poor
quality of the wave function. This value overshoots the real value due
to the fact that the short range wave function, Eq.~(\ref{eq:short})
does not vanish at the origin. For more regular functions such as
those of potential models we expect that logarithmic short distance
enhancement takes place precisely at the fourth and higher orders
where the divergence of the inverse moment becomes manifest. The
chiral TPE deuteron wave functions present the logarithmic enhancement
at fifth order.

\begin{figure}[]
\begin{center}
\epsfig{figure= 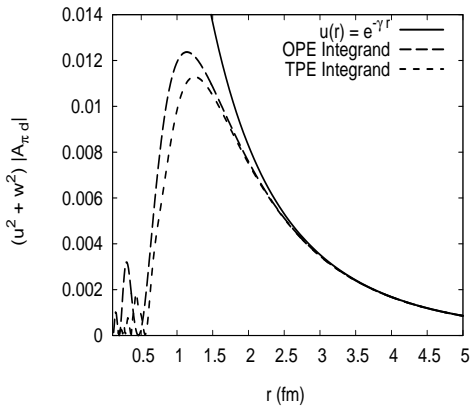,height=5.5cm,width=6.5cm}
\end{center}
\caption{The integrand $(u^2 + w^2) A_{\pi d} (r) $ for the pion-less,
OPE and TPE deuteron wave functions.}
\label{fig:integrand.A_pid.eps}
\end{figure}

In Fig.~\ref{fig:integrand.A_pid.eps} we plot the integrand for a
variety of wave functions. A good feature of the use of the Deloff
function is the {\it irrelevance} of short distance behaviour as
compared to the multiple scattering expansion where there is an
enhancement of short distances. One important lesson we learn from
this exercise is that the appearance of non-analytical enhancements
found in non-perturbative treatments of the NN interaction extends
also to the computation of matrix elements. Strict power counting
simply does not hold, although the corrections are parametrically
small. Numerical results for the pion-less, OPE and TPE potential cases
using the Deloff formula are presented in Table~\ref{tab:a_pid} for
the central values of the $\pi N $ scattering lengths.

As we see, the multiple scattering series provides cancellations which
are {\it independent} on the short distance constraints for the
inverse moments discussed in Sect.~\ref{sec:short}.  One might think
that they might be correlated through the chiral expansion, in the
sense of a perturbative reordering of the truncated multiple
scattering series by re-expanding the convergent inverse moments. This
is unlikely, since the kind of non-analyticities appearing in the
multiple scattering series as a function of $b_0$ and $b_1$ and those
in the inverse moments as a function of the chiral potential parameters
are of quite different nature.

It is worth displaying numerically the convergence of the multiple
scattering series for both the OPE and the TPE deuteron wave
functions. For illustration purposes we have taken the Deloff function
ignoring both binding and recoil corrections (similar features are
observed if these corrections are taken into account). We obtain
\begin{widetext}
\begin{eqnarray}
a_{\pi d}|_{\rm OPE} &=&
\underbrace{130.42\,\tilde{b}_0}_{-0.47013} + 
\underbrace{62.32\,( \tilde{b}_0^2 - 2\,\tilde{b}_1^2 )}_{-2.744} + 
\underbrace{54.84\,
(\tilde{b}_0^3 - 2\,\tilde{b}_0 \tilde{b}_1^2 - 2 \tilde{b}_1^3 )}_{0.367} + 
{\cal O} (b^4 \log b) \\
\nonumber\\
a_{\pi d}|_{\rm TPE} &=&
\underbrace{130.42\,\tilde{b}_0}_{-0.47013} + 
\underbrace{58.25\,( \tilde{b}_0^2 - 2\,\tilde{b}_1^2 )}_{-2.565} + 
\underbrace{37.01\,
( \tilde{b}_0^3 - 2\,\tilde{b}_0 \tilde{b}_1^2 - 2 \tilde{b}_1^3 )}_{0.248} + 
\underbrace{35.97\,
( \tilde{b}_0^4 - 4\,\tilde{b}_0^2 \tilde{b}_1^2 + 2 \tilde{b}_1^4 )}
_{0.035} + 
{\cal O} (b^5 \log b)  \, , 
\label{eq:analytic}
\end{eqnarray}
\end{widetext}
where the results given in the under-braces are in units of
$10^{-2}\,m_{\pi}^{-1}$ (as the pion-deuteron scattering length), and
the $\tilde{b}_i$'s are in fm.  As we see, the bulk of the
contribution is given by the double scattering term. Using the full
OPE and TPE potentials and the full $A_{\pi d} (r) $ operator with no
recoil and binding we get
\begin{eqnarray}
a_{\pi d}^{\rm OPE} = -2.873(3) \cdot 10^{-2} m_{\pi}^{-1} \\ 
a_{\pi d}^{\rm TPE} = -2.77(2) \cdot 10^{-2} m_{\pi}^{-1}  \, , 
\end{eqnarray} 
while the analytic contributions in Eq.~(\ref{eq:analytic}) sum up to
$-2.847$ and $-2.75$ respectively. As we see, the non-analytical
effects are not dramatic, which was not completely obvious {\it a
priori} and to a certain extend are compatible with the uncertainties
in the TPE case. Obviously, very precise estimates might be sensitive to
the non-analytical pieces. In any case, it would be rather interesting
to compute the non-analytical contributions {\it per se}.

\begin{figure}[]
\begin{center}
\epsfig{figure=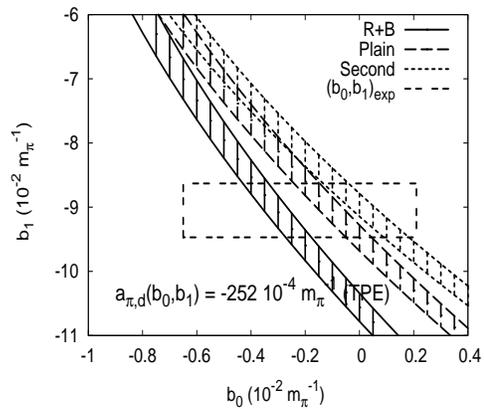,height=5.5cm,width=6.5cm}
\end{center}
\caption{The range of values of the $\pi N$ scattering lengths $b_0$
and $b_1$ in units of $10^{-2} m_\pi^{-1}$ that reproduce the
pion-deuteron scattering length for different approximations (see main
text). The error bars are generated from the experimental error 
in $a_{\pi d}$.
We use the TPE deuteron wave functions with $\eta=0.0256$ and
Set IV~\cite{Entem:2003ft}.} 
\label{fig:b1[b0]}
\end{figure}

Finally, we plot in Fig.~\ref{fig:b1[b0]} the correlation between the
iso-scalar and iso-vector $\pi N$ scattering lengths $b_0 $ and $b_1$
deduced from direct application of the Deloff formula in several
schemes where errors from the $\pi-d $ scattering length and chiral
potential parameters are taken into account. The main source of
uncertainty in all cases turns out to be $a_{\pi d}$. If we take the
iso-vector $b_1$ value from Ref.~\cite{Schroder:1999uq} we obtain in the
TPE case including both binding and recoil corrections the result  
\begin{eqnarray}
b_0 = -0.3(1)  \cdot 10^{-2} m_{\pi}^{-1} \, , 
\end{eqnarray} 
which is compatible with the measured value but about an order of
magnitude more accurate.

\medskip

\medskip

\begin{table}
\begin{center}
\begin{tabular}{|c|c|c|}
\hline \hline
& Static & Recoil \& Binding \\ 
\hline
{\rm $u(r) = e^{-\gamma r}$} & -4.605 & -4.313 \\ 
\hline
{\rm OPE} &  -2.873(3) & -2.446(3) \\ 
{\rm OPE}$^*$ & -2.905(3) & -2.480(2) \\ 
\hline 
{\rm TPE-Set I} & -2.676(11) & -2.238(12) \\ 
{\rm TPE-Set II} & -2.675(12) & -2.237(12) \\ 
{\rm TPE-Set III} & -2.725(16) & -2.29(2) \\ 
{\rm TPE-Set IV} & -2.77(2) & -2.34(2) \\ 
\hline
NijmII & -2.786 & -2.354  \\
Reid93 & -2.789 & -2.357 \\ 
\hline
\hline
\end{tabular}
\end{center}
\caption{ Pion-deuteron scattering length in units of
$10^{-2}\,m_{\pi}^{-1}$ for OPE and TPE potentials in different
approximations. {\it Static} means the Deloff formula without recoil or
binding, and {\it recoil \& binding} taking into account the recoil and 
binding corrections according to Eq.~(\ref{eq:RB-corr}).  
We take $b_0 = -0.22 \cdot 10^{-2}\,m_{\pi}^{-1}$ and $b_1 = -9.05 \cdot
10^{-2}\,m_{\pi}^{-1}$.  OPE and OPE$^*$ account for taking $g_{\pi
NN} = 13.08$ and $g_A = 1.26$ respectively as input. In the OPE case
the error is estimated by varying the semiclassical matching radius in
the range $0.1 - 0.2\,{\rm fm}$, while in the TPE case the error comes
from the experimental uncertainty of the $D/S$ ratio, $\eta =
0.0256(4)$.  TPE Sets I,II,II and IV refer to the chiral parameters,
$c_1$, $c_3$ and $c_4$ of Refs.~\cite{Buettiker:1999ap},
\cite{Rentmeester:1999vw},\cite{Entem:2002sf} and \cite{Entem:2003ft}
respectively.}
\label{tab:a_pid}
\end{table}

\section{Remarks on Boost and finite range corrections}
\label{sec:boost}

One of the effects we have not taken into account in our discussions
has to do with the boost corrections due to the fact that the CM
$\pi-d$ and $\pi-N$ systems do not coincide, so the nucleons inside
the deuteron recoil differently as the deuteron.  Although it has been
argued that these effects are small~\cite{Baru:2004kw} it is
interesting to reanalyze the issue on the light of the present
investigation. Finite range $\pi N $ corrections have also been
computed in Refs.~\cite{Ericson:2000md,Doring:2004kt} and the leading
contribution involves similar operators as in the boost corrections
case, so we will refer mainly to the latter case in the following.
Actually, the effect can be estimated perturbatively yielding an
${\cal O} (Q^4)$ correction according to the modified counting
proposed in Ref.~\cite{Beane:2002wk},
\begin{eqnarray}
a_{\pi-d}|_{\rm Boost} = - \frac1{1 + m_\pi/2M} \frac{m_\pi^2}{8 \pi
M^3 f_\pi^2} ( g_A^2 - 8 M c_2 ) \langle p^2 \rangle \nonumber \\
\end{eqnarray} 
where
\begin{eqnarray}
\langle p^2 \rangle = \int_0^\infty dr \left[ u' (r)^2 + w'(r)^2 +
6 \frac{w(r)^2}{r^2}\right]
\end{eqnarray} 
The operator $\langle p^2 \rangle $ is weakly non-local and strictly
speaking this number is infinite for any singular potential. Actually,
the divergence behaves as $1 / \sqrt{r_c}$ for the OPE potential and
as $1/r_c^3$ for the TPE potential. The large values of $\langle p^2
\rangle $ have been discussed in finite cut-off calculations in
momentum space and attributed to the inner knots of the wave function,
hence favoring the use of relatively small cut-offs. This appears to
be a serious difficulty, but resembles the one faced already for the
inverse moments; one expects non-analytical effects in the couplings
in front of $\langle p^2 \rangle $. This suggests further lines of
research, in particular invoking physically motivated boost
re-summations or including nonlocal effects in the deuteron wave
functions.  As it is known, the higher order contributions to the NN
potential, ${\cal O}(Q^4)$ according to Weinberg's counting, contain
nonlocal pieces. In momentum space and up to NNLO the long distance
part of the potential depends on the momentum transfer $q$ only and
not on the total momentum $k$. Essential non-localities,
i.e. contributions of the form $ V(q,k) = L(q) k^2 $ with $L(q)$ a
non-polynomial function, depend weakly on the total momentum and
appear first at N$^3$LO~\cite{Entem:2002sf} i.e. also ${\cal O} (Q^4)$
due to relativistic $1/M^2$ one loop contributions. In coordinate
space this weak non-locality corresponds to a modification of the
kinetic energy term in the form of a general self-adjoint
Sturm-Liouville operator, $-u''(r) \to - (p(r) u'(r))' $, with a
singular $p(r)$ function at the origin and exponentially decaying at
long distances. It is at present unclear how these non-locally
modified wave functions might influence the pion-deuteron scattering
length and further work along these lines should be pursued. A more
promising perspective consists of resumming boost corrections
non-perturbatively~\footnote{For instance, for the s-wave $\pi N $
term one would have the replacement~\cite{Ericson:1988gk}, $
\frac{m_\pi^2}{M^2} \langle p^2 \rangle \to \Big\langle p^2
\frac{m_\pi^2}{(M+m_\pi)^2-p^2}\Big\rangle $. This corresponds to a
finite nucleon recoil, which has better converging properties.}. Obviously, to achieve a definite conclusion on this issue
one should reexamine the $\pi NN$ system within a full quantum
mechanical Faddeev approach including chiral potentials between the
nucleons which is nontrivial. We leave such a study for future
developments.

\section{Conclusions}
\label{sec:concl} 

In this paper we have analyzed a particular re-summation of re-scattering
effects to pion-deuteron scattering length corresponding to the fixed
center approximation and where recoil and binding effects may be taken
into account to all orders. This re-summation has the important feature
of providing a physically motivated suppression of the deuteron matrix
elements at the origin. This is most welcome since it agrees
qualitatively with the naive expectation that low energy processes
should not depend strongly on short distance details.  The finiteness
of the multiple scattering expansion of the full expression truncated
to a given order requires a series of short distance constraints which
are remarkably fulfilled by the regular solutions of the deuteron
wave functions corresponding to chiral potentials. This is so
precisely because the potentials become singular and are iterated to
all orders. Nevertheless, further insensitivity at short distance can
be gained by using a re-summation formula proposed by Deloff. We have
noted that divergences in multiple scattering expansion arise because
non-analyticities in the chiral expansion appear. This is a rather
general feature which goes beyond just the example of pion-deuteron
scattering addressed in this paper and will also occur in other low
energy reactions. The found non-analyticities are not large
numerically due to the good short distance behaviour of the deuteron
wave functions. Finally, we have compared several ways of
extracting the poorly known iso-scalar $\pi N$ scattering lengths from
the known iso-vector $\pi N $ and the pion-deuteron scattering length,
yielding, as expected, a compatible but more accurate result.

\begin{acknowledgments}

We thank L. Platter, D. R. Phillips, M. D\"oring and A. Deloff for
useful correspondence. We also thank Avraham Gal for pointing out a
mistake in a previous version of the paper.  This research was
supported by DGI and FEDER funds, under contract FIS2004- and by the
Junta de Andaluc\'\i a grant no. FM-225 and EURIDICE grant number
HPRN-CT-2003-00311.

\end{acknowledgments}

\appendix

\section{Short distance contributions of matrix elements involving inverse moments}
\label{sec:app}

In this appendix we quote analytical results for the short distance
contribution to the inverse matrix elements. Most integrals are
sufficiently converging by letting the short distance cut-off to go to
zero. The $1/r^2$ for the OPE case is a bit special since the
integrand presents many oscillations which make numerical
extrapolation unreliable. As pointed out in Ref.~\cite{Platter:2006pt}
it is much better to use the analytical expressions deduced in
Ref.~\cite{PavonValderrama:2005gu}.

\subsection{Short distance OPE}

\begin{figure}[]
\begin{center}
\epsfig{figure=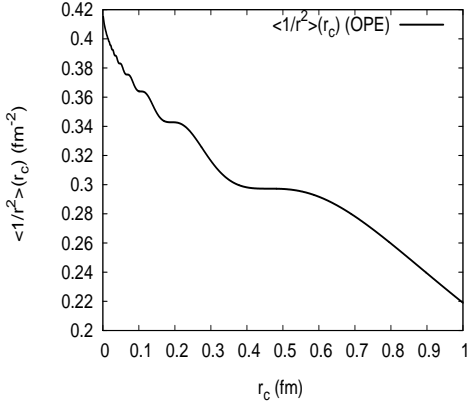,height=5.5cm,width=6.5cm}
\end{center}
\caption{The short distance cut-off dependence (in ${\rm fm}^{-2}$) of
the integrated negative second moment from $r_c$ to infinity,
$\int_{r_c}^\infty (u^2+ w^2) /r^2 dr $ as a function of $r_c$ (in
{\rm fm})}
\label{fig:rm2short[rc]}
\end{figure}

At short distances the OPE potential behaves as 
\begin{eqnarray} 
U_{s}^{\rm OPE} (r) &\to& \frac{R_{s}}{ r^3 } \nonumber \\ 
U_{sd}^{\rm OPE} (r) &\to& \frac{R_{sd}}{ r^3 } \nonumber \\ 
U_{d}^{\rm OPE} (r)  &\to& \frac{R_{d}}{ r^3 } 
\end{eqnarray}
where $R_d = 4 R $, $R_{sd}= 2 \sqrt{2} R $ and $R = 3 g_A^2 M / 32
\pi f_\pi^2 = 1.07764 {\rm fm}$. Going to the diagonal basis the
solution can be written as 
\begin{eqnarray}
u (r) & \to & \sqrt{\frac23} v_A (r)  - \frac{1}{\sqrt{3}} v_R (r) \, , \nonumber  \\
w (r) & \to & \frac1{\sqrt{3}}v_A (r) + \sqrt{\frac23} v_R (r) \, , 
\label{eq:eigenvectors}
\end{eqnarray} 
where 
\begin{eqnarray} 
v_R (r) &= & \left(\frac{r}{R}\right)^{3/4} \left[ C_{1R} e^{+ 4
\sqrt{2} \sqrt{\frac{ R}{r}}} + C_{2R} e^{- 4 \sqrt{2} \sqrt{\frac{
R}{r}}} \right] \, , \nonumber \\ \\ v_A (r) &= &
\left(\frac{r}{R}\right)^{3/4} \left[ C_{1A} e^{- 4 i \sqrt{\frac{
R}{r}}} + C_{2A} e^{ 4 i\sqrt{\frac{ R}{r}}} \right] \, . \nonumber
\label{eq:short_bc}
\end{eqnarray} 
The constants $C_{1R}$, $C_{2R}$, $C_{1A}$ and $C_{2A}$ have been
fixed in Ref.~\cite{PavonValderrama:2005gu} from matching the
numerical solution to a short distance expansion (the regularity
condition $C_{1R}=0$ is imposed). The integral can be computed for
short distances yielding
\begin{widetext}
\begin{eqnarray}
\frac{R^2}{A_S^2 } \int_0^{r_c} dr \frac{u^2+w^2}{r^2} \Big|_{\rm OPE}= 
\frac1{384} \bar
C_{1A}^2 \Big\{ 
&-& 91264\,\pi + 1536\,{\sqrt{x_c}} + 140\,x_c^{\frac{3}{2}}
+ 22816\,{\sqrt{x_c}}\,\cos (\frac{8}{{\sqrt{x_c}}}) \nonumber \\ &-&
1085\,x_c^{\frac{3}{2}}\,\cos (\frac{8}{{\sqrt{x_c}}}) + 2660\,x_c\,\sin
(\frac{8}{{\sqrt{x_c}}}) + 182528\, {\rm Si}(\frac{8}{{\sqrt{x_c}}})
\Big\} \nonumber \\ + \frac1{192} \bar C_{1A} \bar C_{2 A} \Big\{ && 
2660\,x_c\,\cos (\frac{8}{{\sqrt{x_c}}}) + 182528\,{\rm
Ci}(\frac{8}{{\sqrt{x_c}}}) + 31\,{\sqrt{x_c}}\,\left( -736 + 35\,x_c
\right) \,\sin (\frac{8}{{\sqrt{x_c}}}) \Big\} \nonumber \\ + \frac1{384} \bar
C_{2A}^2 \Big\{  && 91264\,\pi + 1536\,{\sqrt{x_c}} + 140\,x_c^{\frac{3}{2}} -
22816\,{\sqrt{x_c}}\,\cos (\frac{8}{{\sqrt{x_c}}}) \nonumber \\ &+&
1085\,x_c^{\frac{3}{2}}\,\cos (\frac{8}{{\sqrt{x_c}}}) - 2660\,x_c\,\sin
(\frac{8}{{\sqrt{x_c}}}) - 182528\,{\rm Si}(\frac{8}{{\sqrt{x_c}}}))
\Big\} + {\cal O} (x_c^2) 
\end{eqnarray} 
\end{widetext} 
with $x_c= r_c /R $  and  Si and Ci are the sine and cosine integral functions
respectively defined as
\begin{eqnarray}
{\rm Si} (z) &=& \int_z^\infty \frac{\sin t}{t} dt \nonumber \\ {\rm
Ci} (z) &=& \int_z^\infty \frac{\sin t}{t} dt
\end{eqnarray} 
The convergence on $r_c$ is shown in Fig.~\ref{fig:rm2short[rc]}. We
see that $0.06 {\rm fm}^{-2}$, i.e.  about $20 \% $ of the result,
comes from the region below $0.2 {\rm fm}$. We have checked that up to
this region the result is rather stable with the given terms. 

\medskip

\subsection{Short distance TPE} 

The short distance behaviour of the TPE has been determined in
Ref.~\cite{Valderrama:2005wv}.  The potential at short distances
behaves as as~\cite{Kaiser:1997mw,Friar:1999sj,Rentmeester:1999vw}
\begin{eqnarray} 
U_{s}^{\rm TPE} (r) &\to& \frac{R_{s}^4}{ r^6 } \nonumber \\ 
U_{sd}^{\rm TPE} (r)  &\to& \frac{R_{sd}^4}{ r^6 } \nonumber \\ 
U_{d}^{\rm TPE} (r)  &\to& \frac{R_{d}}{ r^6 } 
\end{eqnarray} 
where 
\begin{eqnarray} 
(R_{s})^4 &=& \frac{3 g_A^2}{128 f_\pi^4 \pi^2 } ( 4 - 3 g_A^2 + 24 \bar
c_3 - 8 \bar c_4 ) \nonumber \\ (R_{sd})^4 &=& - \frac{3 \sqrt{2}
g_A^2}{128 f_\pi^4 \pi^2 } (-4 + 3 g_A^2 - 16 \bar c_4 ) \nonumber \\
(R_{d})^4 &=& \frac{9 g_A^2}{32 f_\pi^4 \pi^2 } (-1+2 g_A^2 + 2 \bar c_3 -
2 \bar c_4 ) 
\label{eq:vdw_triplet}
\end{eqnarray} 
and $ \bar c_i = M c_i$ are the low energy chiral couplings appearing
in $\pi N $ scattering. 
Diagonalizing the corresponding matrix
\begin{eqnarray} 
\begin{pmatrix}
R_{s}^4 & R_{sd}^4 \\ R_{sd}^4  & R_{d}^4   
\end{pmatrix}  
&=&  
\begin{pmatrix}
\cos\theta & \sin\theta  \\ -\sin\theta &  \cos \theta   
\end{pmatrix}  
\begin{pmatrix}
-R_{+}^4  & 0  \\ 0 &  -R_{-}^4     
\end{pmatrix} \nonumber \\ &\times&   
\begin{pmatrix}
\cos\theta & -\sin\theta  \\ \sin\theta &  \cos \theta   
\end{pmatrix} 
\end{eqnarray}  
Note that the potential is negative definite. In the diagonal basis one has 
\begin{eqnarray} 
\begin{pmatrix}
u \\ w 
\end{pmatrix}  
&      \to & 
\begin{pmatrix}
\cos\theta & \sin\theta  \\ -\sin\theta &  \cos \theta   
\end{pmatrix}  
\begin{pmatrix}
v_+ \\ v_- 
\end{pmatrix}  
\end{eqnarray}  
where the short distance eigen functions are 
\begin{eqnarray} 
v_+ (r) &=& \left(\frac{r}{R_+}\right)^{\frac32} \left\{ C_{+,s}
\sin\left[ \frac{1}{2} \frac{R_+^2}{r^2} \right] + C_{+,c} \cos\left[
\frac{1}{2} \frac{R_+^2}{r^2} \right] \right\} \nonumber \\ v_- (r)
&=& \left(\frac{r}{R_-}\right)^{\frac32} \left\{ C_{-,s} \sin\left[
\frac{1}{2} \frac{R_-^2}{r^2} \right] + C_{-,c} \cos\left[ \frac{1}{2}
\frac{R_-^2}{r^2} \right] \right\} \nonumber \\ 
\end{eqnarray}  

\medskip
\begin{figure}[ttt]
\begin{center}
\epsfig{figure=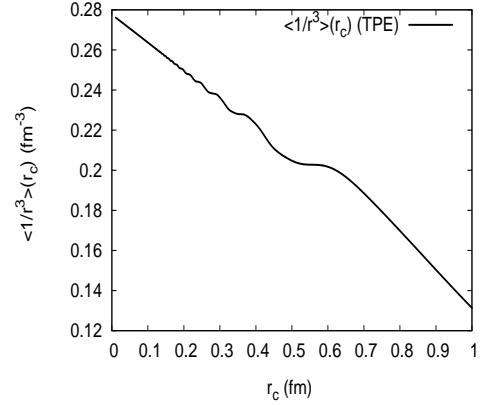,height=5.5cm,width=6.5cm}
\end{center}
\caption{The short distance cut-off dependence of (in ${\rm fm}^{-3}$) the
integrated negative second moment from $r_c$ to infinity,
$\int_{r_c}^\infty (u^2+ w^2) /r^3 dr $ in the TPE case as a function of
$r_c$ (in {\rm fm}). We use Set IV of chiral constants.}
\label{fig:rm3short[rc]}
\end{figure}
Higher order contributions could in principle be computed similarly to
what was done in the OPE case~ \cite{PavonValderrama:2005gu} although
such refinements will not be needed here. Matching our numerical
solutions to these short distance solutions at $r_c=0.1 {\rm fm}$ we
get (for Set IV of parameters) $R_+ = 2.11 {\rm fm}$ and $R_- =1.16
{\rm fm}$
\begin{eqnarray}
 C_{+,c} &=& 6.174 - 257.329 \eta \nonumber \\ 
C_{+,s}  &=& -1.585 + 38.745 \eta \nonumber \\ 
C_{-,c}  &=& 2.451 - 118.84 \eta   \nonumber \\ 
C_{-,s}  &=& 2.522 - 84.40 \eta 
\end{eqnarray} 
for the long distance normalization such that $u (r) \to e^{-\gamma r}
$.  The integral for the inverse moment $\langle r^{-3} \rangle $ can
be computed analytically using that 
\begin{eqnarray}
\int_0^{r_c} \frac{u^2+ w^2}{r^n} dr = \int_0^{r_c} \frac{v_+^2+
v_-^2}{r^n} dr
\end{eqnarray} 
and the appropriate normalization. 
For the third inverse moments the integral can be written
in terms of Fresnel integrals. The result is depicted in
Fig.~\ref{fig:rm3short[rc]}. As we see, the short distance
contribution for the moments is rather small. The contributions from  
the region between the origin and $r_c=0.2 {\rm fm}$ are given by
\begin{eqnarray}
\langle r^{-1} \rangle_{\rm TPE}^{\rm short} &=& 0.0004 {\rm fm}^{-1}
\nonumber \\ \langle r^{-2} \rangle_{\rm TPE}^{\rm short} &=& 0.003
{\rm fm}^{-2} \nonumber \\ \langle r^{-3} \rangle_{\rm TPE}^{\rm
short} &=& 0.025 {\rm fm}^{-3}\nonumber \\
\end{eqnarray}


\end{document}